\documentclass[]{aastex631}
\usepackage{amsmath, graphicx, subfigure, comment, bm, amssymb, scalerel}
\usepackage{esvect, epstopdf, CJK}

\newcommand{\kepler}{{Kepler}}
\newcommand{\corot}{{CoRoT}}

\newcommand{\numax}{\mbox{$\nu_{\rm max}$}}
\newcommand{\Dnu}{\mbox{$\Delta \nu$}}

\newcommand{\muHz}{\mbox{$\mu$Hz}}

\newcommand{\msun}{\!{M_{\sun}}}

\shorttitle{Asteroseismic analysis of HD\,76920}
\shortauthors{Jiang et al.}

\begin{document}
\begin{CJK*}{UTF8}{gbsn}

\title{TESS Asteroseismic Analysis of HD\,76920: The Giant Star Hosting An Extremely Eccentric Exoplanet}

\correspondingauthor{Chen Jiang}
\email{jiangc@mps.mpg.de}
\correspondingauthor{Tao Wu}
\email{wutao@ynao.ac.cn}

\author[0000-0002-7614-1665]{Chen Jiang(姜晨)}
\affiliation{Max-Planck-Institut f{\"u}r Sonnensystemforschung, Justus-von-Liebig-Weg 3, 37077 G{\"o}ttingen, Germany}

\author[0000-0001-6832-4325]{Tao Wu (吴涛)}
\affiliation{Yunnan Observatories, Chinese Academy of Sciences, 396 Yangfangwang, Guandu District, Kunming, 650216, People's Republic of China}
\affiliation{Key Laboratory for the Structure and Evolution of Celestial Objects, Chinese Academy of Sciences, 396 Yangfangwang, Guandu District, Kunming, 650216, People's Republic of China}
\affiliation{Center for Astronomical Mega-Science, Chinese Academy of Sciences, 20A Datun Road, Chaoyang District, Beijing, 100012, People's Republic of China}
\affiliation{University of Chinese Academy of Sciences, Beijing 100049, People's Republic of China}
\affiliation{Institute of Theoretical Physics, Shanxi University, Taiyuan 030006, China}

\author[0000-0002-9464-8101]{Adina~D.~Feinstein}
\altaffiliation{NSF Graduate Research Fellow}
\affiliation{Department of Astronomy and Astrophysics, University of Chicago, 5640 S. Ellis Ave, Chicago, IL 60637, USA}

\author[0000-0002-3481-9052]{Keivan G.\ Stassun}
\affiliation{Department of Physics and Astronomy, Vanderbilt University, Nashville, TN 37235, USA}

\author[0000-0001-5222-4661]{Timothy R. Bedding}
\affiliation{Sydney Institute for Astronomy (SIfA), School of Physics, University of Sydney, NSW 2006, Australia}
\affiliation{Stellar Astrophysics Centre (SAC), Department of Physics and Astronomy, Aarhus University, Ny Munkegade 120, 8000 Aarhus C, Denmark}

\author[0000-0001-8014-6162]{Dimitri Veras}
\affiliation{Centre for Exoplanets and Habitability, University of Warwick, Coventry CV4 7AL, UK}
\affiliation{Centre for Space Domain Awareness, University of Warwick, Coventry CV4 7AL, UK}
\affiliation{Department of Physics, University of Warwick, Coventry CV4 7AL, UK}

\author[0000-0001-8835-2075]{Enrico Corsaro}
\affiliation{INAF --- Osservatorio Astrofisico di Catania, via S.~Sofia 78, 95123 Catania, Italy}

\author[0000-0002-1988-143X]{Derek L. Buzasi}
\affiliation{Department of Chemistry \& Physics, Florida Gulf Coast University, 10501 FGCU Boulevard S., Fort Myers, FL 33965, USA}

\author[0000-0002-4879-3519]{Dennis Stello}
\affiliation{School of Physics, University of New South Wales, NSW 2052, Australia}
\affiliation{Sydney Institute for Astronomy (SIfA), School of Physics, University of Sydney, NSW 2006, Australia}
\affiliation{Stellar Astrophysics Centre (SAC), Department of Physics and Astronomy, Aarhus University, Ny Munkegade 120, 8000 Aarhus C, Denmark}

\author[0000-0003-3020-4437]{Yaguang Li(李亚光)}
\affiliation{Sydney Institute for Astronomy (SIfA), School of Physics, University of Sydney, NSW 2006, Australia}
\affiliation{Stellar Astrophysics Centre (SAC), Department of Physics and Astronomy, Aarhus University, Ny Munkegade 120, 8000 Aarhus C, Denmark}

\author[0000-0002-0129-0316]{Savita Mathur}
\affil{Instituto de Astrof\'isica de Canarias (IAC), E-38205 La Laguna, Tenerife, Spain}
\affil{Universidad de La Laguna (ULL), Departamento de Astrof\'isica, E-38206 La Laguna, Tenerife, Spain}

\author[0000-0002-8854-3776]{Rafael A. Garc\'{i}a}
\affil{Universit\'e Paris-Saclay, Universit\'e Paris Cit\'e, CEA, CNRS, AIM, 91191, Gif-sur-Yvette, France}

\author[0000-0003-0377-0740]{Sylvain N. Breton}
\affiliation{D\'epartement d'Astrophysique/AIM, CEA/IRFU, CNRS/INSU, Univ. Paris-Saclay \& Univ. de Paris, 91191 Gif-sur-Yvette, France}

\author[0000-0002-8661-2571]{Mia S. Lundkvist}
\affiliation{Stellar Astrophysics Centre (SAC), Department of Physics and Astronomy, Aarhus University, Ny Munkegade 120, DK-8000 Aarhus C, Denmark}

\author[0000-0001-8916-8050]{Przemys\l{}aw J. Miko\l{}ajczyk}
\affiliation{Astronomical Observatory, University of Warsaw, Al. Ujazdowskie 4, 00-478 Warsaw, Poland}
\affiliation{Astronomical Institute, University of Wroc\l{}aw, Miko\l{}aja Kopernika 11, 51-622 Wroc\l{}aw, Poland}

\author[0000-0002-0833-7084]{Charlotte Gehan}
\affiliation{Max-Planck-Institut f{\"u}r Sonnensystemforschung, Justus-von-Liebig-Weg 3, 37077 G{\"o}ttingen, Germany}
\affiliation{Instituto de Astrof\'{\i}sica e Ci\^{e}ncias do Espa\c{c}o, Universidade do Porto,  Rua das Estrelas, 4150-762 Porto, Portugal}

\author[0000-0002-4588-5389]{Tiago L. Campante}
\affiliation{Instituto de Astrof\'{\i}sica e Ci\^{e}ncias do Espa\c{c}o, Universidade do Porto,  Rua das Estrelas, 4150-762 Porto, Portugal}
\affiliation{Departamento de F\'{\i}sica e Astronomia, Faculdade de Ci\^{e}ncias da Universidade do Porto, Rua do Campo Alegre, s/n, 4169-007 Porto, Portugal}

\author[0000-0002-9480-8400]{Diego Bossini}
\affiliation{Instituto de Astrof\'{\i}sica e Ci\^{e}ncias do Espa\c{c}o, Universidade do Porto,  Rua das Estrelas, 4150-762 Porto, Portugal}

\author[0000-0002-7084-0529]{Stephen R. Kane}
\affiliation{Department of Earth and Planetary Sciences, University of California, Riverside, CA 92521, USA}

\author[0000-0001-7664-648X]{Jia Mian Joel Ong (王加冕)}
\affiliation{Department of Astronomy, Yale University, P.O.~Box 208101, New Haven, CT 06520-8101, USA}

\author[0000-0002-7772-7641]{Mutlu Y{\i}ld{\i}z}
\affiliation{Department of Astronomy and Space Sciences, Science Faculty, Ege
University, 35100, Bornova, \.Izmir, T{\"u}rkiye}

\author[0000-0001-9198-2289]{Cenk Kayhan}
\affiliation{Department of Astronomy and Space Sciences, Science Faculty, Erciyes University, 38030, Melikgazi, Kayseri, T{\"u}rkiye}

\author[0000-0002-9424-2339]{Zeynep \c{C}el\.{i}k Orhan}
\affiliation{Department of Astronomy and Space Sciences, Science Faculty, Ege
University, 35100, Bornova, \.Izmir, T{\"u}rkiye}

\author[0000-0001-5759-7790]{S\.{i}bel \"Ortel}
\affiliation{Department of Astronomy and Space Sciences, Science Faculty, Ege
University, 35100, Bornova, \.Izmir, T{\"u}rkiye}

\author[0000-0003-1860-1851]{Xinyi Zhang (张昕旖)}
\affiliation{State Key Laboratory of Lunar and Planetary Sciences, Macau University of Science and Technology, Macau, People’s Republic of China}

\author[0000-0001-8237-7343]{Margarida S. Cunha}
\affiliation{Instituto de Astrof\'{\i}sica e Ci\^{e}ncias do Espa\c{c}o, Universidade do Porto,  Rua das Estrelas, 4150-762 Porto, Portugal}
\affiliation{Departamento de F\'{\i}sica e Astronomia, Faculdade de Ci\^{e}ncias da Universidade do Porto, Rua do Campo Alegre, s/n, 4169-007 Porto, Portugal}

\author[0000-0001-6295-3526]{Bruno Lustosa de Moura}
\affiliation{Departamento de Fisica, Universidade Federal do Rio Grande do Norte, 59072-970 Natal, RN, Brazil}
\affiliation{Instituto Federal do Rio Grande do Norte—IFRN, Brazil}

\author[0000-0002-0007-6211]{Jie Yu (余杰)}
\affiliation{Max-Planck-Institut f{\"u}r Sonnensystemforschung, Justus-von-Liebig-Weg 3, 37077 G{\"o}ttingen, Germany}

\author[0000-0001-8832-4488]{Daniel Huber}
\affiliation{Institute for Astronomy, University of Hawai`i, 2680 Woodlawn Drive, Honolulu, HI 96822, USA}


\author[0000-0002-6176-7745]{Jian-wen, Ou (欧建文)}
\affiliation{School of Physics and Electromechanical Engineering, Shaoguan University, 512005 Shaoguan, Guangdong Province, China}

\author[0000-0001-9957-9304]{Robert A. Wittenmyer}
\affiliation{University of Southern Queensland, Centre for Astrophysics, USQ Toowoomba, QLD 4350, Australia}

\author[0000-0001-7696-8665]{Laurent Gizon}
\affiliation{Max-Planck-Institut f{\"u}r Sonnensystemforschung, Justus-von-Liebig-Weg 3, 37077 G{\"o}ttingen, Germany}
\affiliation{Institut für Astrophysik, Georg-August-Universit\"at G\"ottingen,  Friedrich-Hund-Platz 1, 37077 G\"ottingen, Germany}
\affiliation{Center for Space Science, NYUAD Institute, New York University Abu Dhabi, Abu Dhabi, UAE}

\author[0000-0002-5714-8618]{William J. Chaplin}
\affiliation{Stellar Astrophysics Centre (SAC), Department of Physics and Astronomy, Aarhus University, Ny Munkegade 120, 8000 Aarhus C, Denmark}
\affiliation{School of Physics and Astronomy, University of Birmingham, Birmingham B15 2TT, UK}

\begin{abstract}

The Transiting Exoplanet Survey Satellite (TESS) mission searches for new exoplanets. The observing strategy of TESS results in high-precision photometry of millions of stars across the sky, allowing for detailed asteroseismic studies of individual systems.
In this work, we present a detailed asteroseismic analysis of the giant star HD\,76920 hosting a highly eccentric giant planet ($e = 0.878$) with an orbital period of 415 days, using 5 sectors of TESS light curve that cover around 140 days of data. Solar-like oscillations in HD\,76920 are detected around  $52 \, \muHz$ by TESS for the first time. By utilizing asteroseismic modeling that takes classical observational parameters and  stellar oscillation frequencies as constraints, we determine improved measurements of the stellar mass ($1.22 \pm 0.11\, \msun$), radius ($8.68 \pm 0.34\,R_\sun$), and age ($5.2 \pm 1.4\,$Gyr). With the updated parameters of the host star, we update the semi-major axis and mass of the planet as $a=1.165 \pm 0.035$\,au and $M_{\rm p}\sin{i} = 3.57 \pm 0.22\,M_{\rm Jup}$. With an orbital pericenter of $0.142 \pm 0.005$\,au, we confirm that the planet is currently far away enough from the star to experience negligible tidal decay until being engulfed in the stellar envelope. We also confirm that this event will occur within about 100\,Myr, depending on the stellar model used.
\end{abstract}


\section{Introduction} \label{sec:intro}

The excellent quality of photometric data from space observation missions, such as \corot\ \citep{baglin06} and \kepler\ \citep{borucki10}, allow for 
major advancements in the understanding of stellar interior physics and evolution using asteroseismology. Asteroseismology is the study of internal structure of stars by the interpretation of their oscillation frequencies. In particular, the detection of oscillations in solar-type and red giant stars has led to breakthroughs such as the discovery of fast core rotation \citep{beck12} and a way to distinguish between hydrogen-shell-burning stars and stars that are also burning helium in their cores \citep{bedding11}. 

The advent of space photometry has also brought in advancements of data analysis techniques \citep[e.g.,][]{davies16, lund17, corsaro2014, corsaro15} and stellar modeling strategies \citep[e.g.,][]{wu2016, seren17, silva17, wu2017}. Furthermore, by using individual oscillation frequencies as constraints in the model optimization process \citep{metcalfe10, jiang2011, mathur12, paxton13, rendle19}, asteroseismic modeling has proven to be a robust tool to determine fundamental stellar properties, including stellar distances \citep{silva12,rodrigues14}, radii and masses \citep{casa14,pinson14,sharma16}, and most importantly ages and core size for red giants and clump stars \citep{casa16,anders17,pinson18, zhang2018, zhang2020}. Consequently, this enables us to characterize systematically the properties of the exoplanet-host stars through asteroseismology, which in turn provides an unprecedented level of precision in the parameters estimated for the planets \citep{ballard14,camp15,lundkvist16,kayhan2019}. Furthermore, the synergy between asteroseismology and exoplanetary research also enables us to set constraints on the spin-orbit alignment of exoplanet systems \citep{huber13, benomar14, chaplin14, lund14, camp16a, kamiaka19} and to perform statistical inferences on the planetary orbital eccentricities, by making use of asterodensity profiling \citep{kane12,sliski14,van15,van19}.

The Transiting Exoplanet Survey Satellite (TESS) Mission \citep{ricker15} is NASA's near all-sky survey for exoplanets, which launched in 2018. The large sample of monitored systems guarantees the synergy between asterosismology and exoplanetary science to continue to expand  \citep{camp18,huber18,hatt2022}.
TESS searches for exoplanets using the transit method in an area 400 times larger than that covered by the \kepler\ mission.
Although the exploration of new exoplanetary systems is the main scientific goal of the mission, TESS has also provided aids toward the characterization of previously known systems \citep{kane2021}.
Thanks to the high-quality of TESS photometry and large sky coverage, oscillations are expected to be detected in hundreds of thousands of solar-like oscillators \citep{camp18,huber18, schofield19}, including several hundred asteroseismic exoplanet hosts \citep{camp16b}. 
Detections of oscillations by TESS in previously known exoplanet-host stars were reported by several works \citep[e.g.][]{camp19, jiang2020a, nielsen20, hill21, huber2022}, following on the discovery of the first planet transiting a star (TOI-197) in which oscillations could be measured by TESS \citep{huber19}. These extraordinary synergies between asteroseismology and exoplanetary science significantly improve our understanding of planet systems outside of the solar system and provide insight into the occurrence rates of exoplanets as a function of their host stars' property and evolutionary state, as well as the planets' mass, size and orbital architecture. One of the most interesting architectures is the planets having very eccentric orbits that are considered as a possible origin of hot Jupiter \citep[see][for a review]{dawson2018}. Planets in very eccentric orbits are a prime example and testbed of how planetary systems form and evolve. 

The first planet discovered to orbit around a evolved giant star, $\iota$\,Draconis\,b, is on a highly eccentric orbit with $e = 0.71$ \citep{frink2002, kane2010}. Even larger orbital eccentricity is found in a planet around the giant star HIP\,126844 with $e=0.76$ \citep{adamow2012, adamow2018}.
In this work, using TESS data, we present an asteroseismic analysis of the K giant star HD\,76920 (TIC\,302372658) known to host a planet. At the time that HD\,76920\,b was first detected through the radial-velocity (RV) survey of Pan-Pacific Planet Search \citep{wittenmyer17}, it was the most eccentric exoplanet known to orbit an evolved star, with an orbital eccentricity of $0.856\pm0.009$. Later, with the help of new multi-site RV measurements, \cite{bergmann21} refined the planetary properties, finding an even higher eccentricity of $0.8783 \pm 0.0025$ and an orbital period of $415.891^{+0.043}_{-0.039}$ days, a minimum planet mass of $3.13^{+0.41}_{-0.43} \,M_{\rm Jup}$, and a semi-major axis of $1.091^{+0.068}_{-0.077}$\,au. 
There is no evidence of any unseen binary companion, suggesting a scattering event rather than Kozai oscillations as the probable explanation for the observed eccentricity, and making the system valuable to the study the evolution and occurrence of planets around evolved stars. TESS detected solar-like oscillations in HD\,76920 for the first time. \cite{bergmann21} analyzed three sectors (9, 10 and 11) of TESS data and estimated the stellar mass and radius through the scaling relations \citep{brown1991, kjeldsen1995, stello2008, kallinger2010} using the measurements of global seismic parameters. While the scaling relations provide decent estimates of the stellar mass and radius for stars showing solar-like oscillations, improvements, in terms of accuracy and precision, of the estimates can be achieved by using asteroseismic modeling and individual oscillation modes.

In this paper, we aim to analyze the solar-like oscillations from 5 sectors of TESS photometric light curve. Including these oscillations in detailed stellar modeling can help us derive precise fundamental stellar properties of the host star. 
The improved stellar parameters can further be used to update the properties of the orbiting planet, which is of great importance in characterizing the planetary system. 
The determined stellar age from asteroseismology and modeling provide information about the evolution of the system, which assists at predicting the final fate of the planet. 

\begin{table}
\begin{center}
\caption{Stellar Parameters for HD\,76920}
\label{tb:starpar}
\renewcommand{\tabcolsep}{0mm}
\begin{tabular}{l c c}
\noalign{\smallskip}
\tableline\tableline
\noalign{\smallskip}
\textbf{Parameter} & \textbf{Value} & \textbf{References} \\
\noalign{\smallskip}
\tableline
\noalign{\smallskip}
\multicolumn{3}{c}{Basic Properties} \\
\noalign{\smallskip}
\hline
\noalign{\smallskip}
TIC & 302372658 & 1 \\
\textit{Hipparcos} ID & 43803 & 2 \\
TESS Mag. & 6.87 & 1 \\
Sp.~Type & K1 III & 3 \\
\noalign{\smallskip}
\hline
\noalign{\smallskip}
\multicolumn{3}{c}{Spectroscopy} \\
\noalign{\smallskip}
\hline
\noalign{\smallskip}
$T_{\rm eff}$ (K) & $4698 \pm 100$ & 4 \\
& $4664 \pm 53$ & 5\\
$[{\rm Fe}/{\rm H}]$ (dex) & $-0.11 \pm 0.10$ & 4 \\
& $-0.19 \pm 0.06$ & 5 \\
$\log g$ (cgs) & $2.94 \pm 0.15$ & 4 \\
& $2.71 \pm 0.04$ & 5\\
\noalign{\smallskip}
\hline
\noalign{\smallskip}
\multicolumn{3}{c}{SED \& Gaia DR3 Parallax\tablenotemark{a} }\\
\noalign{\smallskip}
\hline
\noalign{\smallskip}
$\pi$ (mas) & $5.4618 \pm 0.0187$ & 6 \\
$A_V$ & $0.20 \pm 0.05$ & 7 \\
& $0.15 \pm 0.07$ & 7 \\
$F_{\rm bol}$ (${\rm erg\,s^{-1}\,cm^{-2}}$) & $(3.12 \pm 0.11) \times 10^{-8}$ & 7 \\
& $(3.08 \pm 0.11) \times 10^{-8}$ & 7 \\
$R_\star$ ($R_\sun$) & $8.64 \pm 0.37$ & 7 \\
& $8.70 \pm 0.25$ & 7 \\
$L_\star$ ($L_\sun$) & $32.54 \pm 1.17$ & 7 \\
& $32.06 \pm 1.16$ & 7 \\
$M_\star$ ($\msun$) & $1.68 \pm 0.10$\tablenotemark{b} & 7 \\
\noalign{\smallskip}
\hline
\noalign{\smallskip}
\multicolumn{3}{c}{Asteroseismology} \\
\noalign{\smallskip}
\hline
\noalign{\smallskip}
$\Delta\nu$ ($\mu$Hz) & $5.53 \pm 0.15$ & 7 \\
$\nu_{\rm max}$ ($\mu$Hz) & $53.2 \pm 2.3$ & 7 \\
$M_{\star, \rm seis}$ ($\msun$) & $1.22 \pm 0.11$ & 7 \\
$R_{\star, \rm seis}$ ($R_\sun$) & $8.68 \pm 0.34$  & 7 \\
$\rho$ (gcc) & $0.0026 \pm 0.0004$  & 7 \\
$\log g$ (cgs) & $2.648 \pm 0.037$  & 7 \\
$t$ (Gyr) & $5.2 \pm 1.4$  & 7 \\
\noalign{\smallskip}
\hline
\noalign{\smallskip}
\end{tabular}
\end{center}
\tablenotetext{a}{\scriptsize  Two SED pipelines are used to obtain $A_V$, $F_{\rm bol}$, $R_\star$ and $L_\star$ (Section~\ref{sc:spectro}). The second value of each parameter is from the SEDEX pipeline \citep{yu2021, yu2022}.}
\tablenotetext{b}{\scriptsize Based on extrapolated relations of \citet{torres10}.}
\tablerefs{\scriptsize (1) \citet{stass18b}, (2) \citet{HIP}, (3) \citet{houk75}, (4) \citet{wittenmyer17}, (5) \citet{bergmann21}, (6) \citet{GaiaEDR3}, (7) this work.}
\end{table}

\section{Observations}

\subsection{TESS Observations}
\label{sc:timeseries}

\begin{figure}[!ht]
\begin{center}
\resizebox{1.0\hsize}{!}{\includegraphics{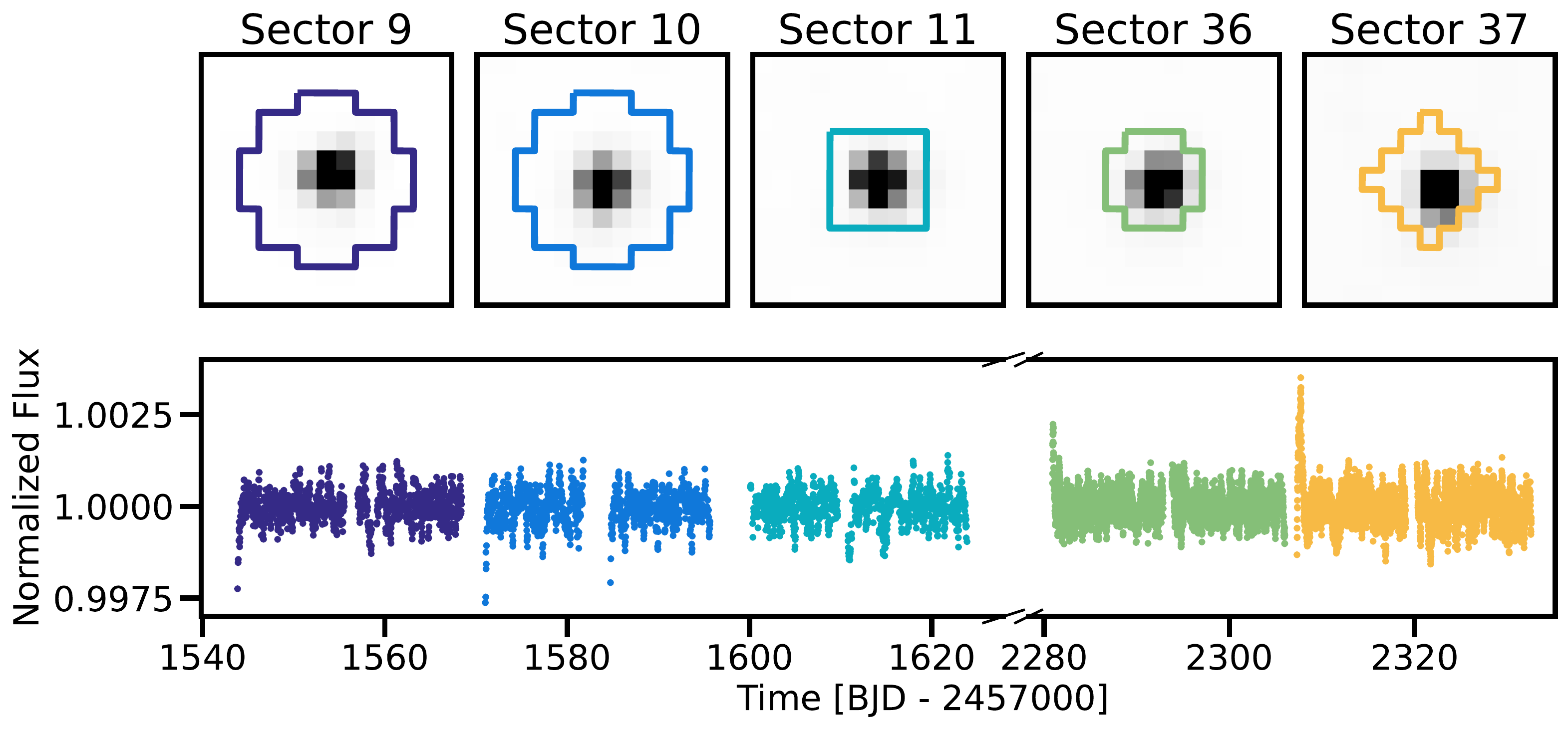}}
\caption{\texttt{eleanor} best-fit apertures (top) overlaid on the TESS target pixel files (TPFs) extracted per each sector. The TPFs are all scaled from 0 to 30000 e$^{-1}$ s$^{-1}$. We use these apertures to extract the flux within \texttt{eleanor}, which are then corrected via the default \texttt{eleanor} corrected flux routine (bottom). Apertures and light curves are colored by TESS sector. \label{fg:eleanor}}
\end{center}
\end{figure}

HD\,76920 has been observed by TESS in Sectors 9, 10, 11 at 30-min cadence during its Cycle 1 observations, and Sectors 36 and 37 at 10-min cadence\footnote{HD\,76920 has also been observed in Sector 38 at 10-min cadence, but the data was not released at the time of this analysis.} during its Cycle 3 observations. To extract these light curves, we used the open-source Python package \texttt{eleanor} \citep[v2.0.3; ][]{2019PASP..131i4502F}\footnote{For sectors 36 and 37, \texttt{eleanor} uses the \texttt{TessCut} tool \citep{2019ascl.soft05007B}.}. \texttt{eleanor} performs background subtraction, systematics corrections, and aperture selection per each sector of data. We extracted $13 \times 13$ pixels postage stamps and applied the \texttt{eleanor} default apertures for light curve extraction (top panel of Figure~\ref{fg:eleanor}). Although the aperture selection process is optimized for exoplanet searches, we found also these default apertures to perform well for asteroseismic measurements. 

We used the \texttt{eleanor} corrected flux; this flux option corrects for systematic issues by regressing against a linear model of time, background, and position and removing trends by the co-trending basis vectors provided by the Science Process Operations Center pipeline \citep{2016SPIE.9913E..3EJ}. We additionally applied quality masks, provided by \texttt{eleanor}, to remove any bad cadences and removed outliers $\geq 7 \sigma$ from the median of each light curve. The resulting light curves are shown in the bottom panel of Figure~\ref{fg:eleanor}. No transit of the planet was detected in existing TESS data due to the short observation coverage compared to its long orbital period. TESS is expected to observe HD\,76920 again for 5 months in Cycle 5. 

\subsection{Broadband Photometry and Spectral Energy Distribution}
\label{sc:spectro}

\begin{figure}
\resizebox{1.0\hsize}{!}{\includegraphics[angle=90]{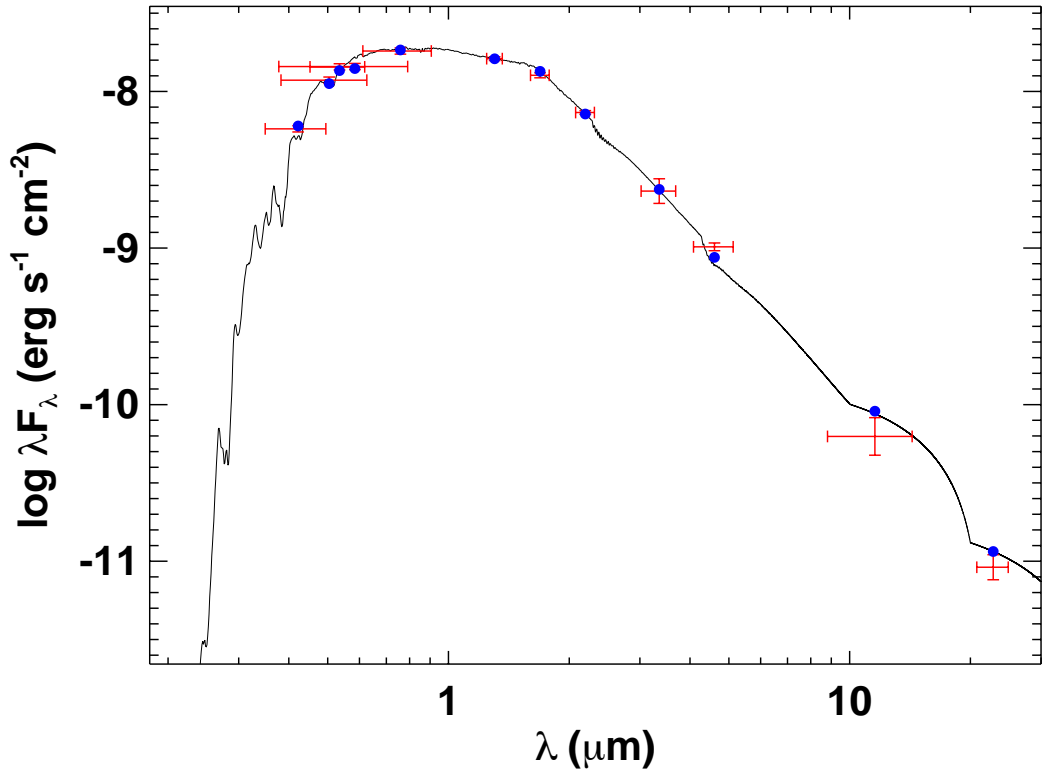}}
\caption{Spectral energy distribution of HD 76920. Red symbols represent the observed photometric measurements, where the horizontal bars represent the effective width of the passband. Blue symbols are the model fluxes from the best-fit Kurucz atmosphere model (black). }
\label{fg:sed}
\end{figure}

As an independent determination of the basic stellar parameters, we performed an analysis of the broadband spectral energy distribution (SED) of the star together with the Gaia DR3 parallax \citep[with no systematic offset applied; see, e.g.,][]{stassun21}, in order to determine an empirical measurement of the stellar radius, following the procedures described in \cite{stassun16} and \cite{stassun17,stassun18}. We pulled the $B_T V_T$ magnitudes from Tycho-2, the $JHK_S$ magnitudes from the Two Micron All Sky Survey, the W1--W4 magnitudes from the Wide-field Infrared Survey Explorer, and the $G_{\rm BP} G_{\rm RP}$ magnitudes from Gaia DR3. Together, the available photometry spans the full stellar SED over the wavelength range 0.4--22\,$\mu$m (see Figure~\ref{fg:sed}).  

We performed a fit using Kurucz stellar atmosphere models, with the effective temperature ($T_{\rm eff}$), surface gravity ($\log g$), and metallicity ([Fe/H]) adopted from the spectroscopic analysis of \cite{wittenmyer17}. The remaining free parameter is the extinction $A_V$, which we limited to the maximum line-of-sight extinction from the Galactic dust maps of \citet{schlegel98}. The resulting fit (Figure~\ref{fg:sed}) has a reduced $\chi^2$ of 2.2, and best-fit $A_V = 0.20 \pm 0.05$. Integrating the (unreddened) model SED gives the bolometric flux at Earth, $F_{\rm bol} = 3.12 \pm 0.11 \times 10^{-8}$ erg~s$^{-1}$~cm$^{-2}$. Taking the $F_{\rm bol}$ and $T_{\rm eff}$ together with the Gaia parallax, gives the stellar radius, $R_\star = 8.64 \pm 0.37\,R_\sun$, revealing the star to be clearly evolved. In addition, we can estimate the stellar mass directly from $R_\star$ together with the spectroscopic $\log g$ from \cite[][see Table 1]{wittenmyer17}, which gives $M_\star = 2.37 \pm 0.41$~M$_\odot$; this is higher than the value obtained from the empirical relations of \citet{torres10}, giving $M_\star = 1.68 \pm 0.10$~M$_\odot$, indicating an overestimate of spectroscopic $\log g$ (see Section~\ref{sc:model}). Combining $F_{\rm bol}$ and Gaia DR3 parallax allows us to derive a luminosity $L_\star = 32.54 \pm 1.17\, L_\sun$.
We also performed an independent SED fitting using the SEDEX pipeline \citep{yu2021, yu2022} with MARCS atmosphere models \citep{gustafsson2008} and same input atmospheric parameters. The resulting parameters from both SED fittings are well consistent, as listed in Table~\ref{tb:starpar}.

\section{Asteroseismic Analysis}
\label{sc:asteroseismic}

\subsection{Global Oscillation Parameters}
\label{sc:global}

\begin{figure}
\resizebox{1.0\hsize}{!}{\includegraphics[]{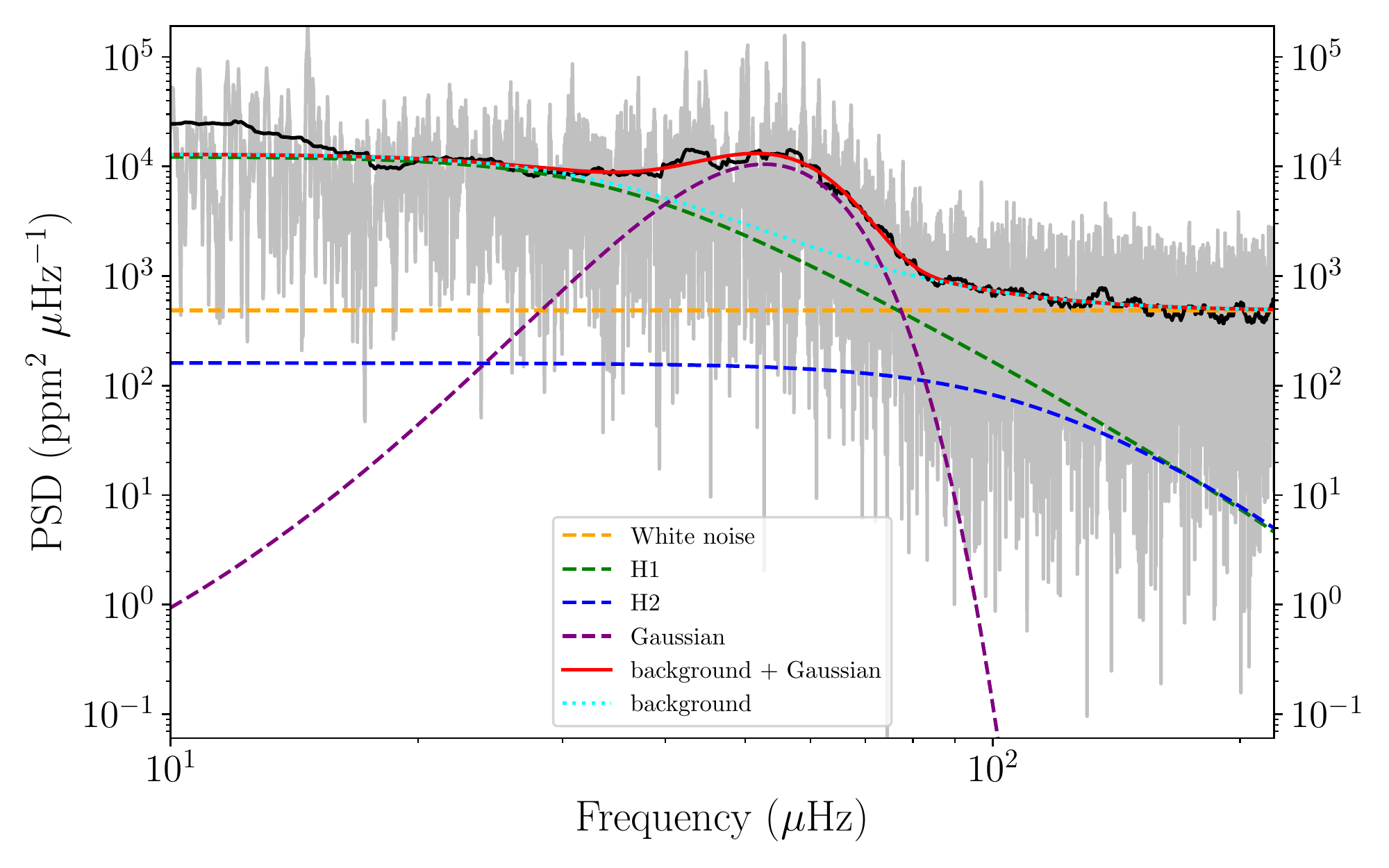}}
\caption{Power spectral density (PSD) of HD\,76920 and corresponding global background model fit (cyan dotted curve). The background model consist of two Harvey-like profiles (green and blue dashed curves) and white noise (orange dashed line). The solid red curve depicts the summation of the background and a Gaussian fit (purple dashed) to the oscillation power excess envelope.
The PSD is generated using 5 sectors of TESS data (Figure~\ref{fg:eleanor}). The 10-min cadence sectors are rebinned to 30-min cadence, with simple average over three measurements. The original PSD is shown in gray and a heavily smoothed (Gaussian with an FWHM of $\Dnu$) version in black. The fit displayed here disregarded frequencies below $\sim$10\,$\muHz$ (Section~\ref{sc:global}).}
\label{fg:powerspectrum}
\end{figure}

For the seismic analysis, the TESS light curve prepared by \texttt{eleanor} was distributed to several groups using different pipelines \citep[e.g.,][]{huber2009, mathur2010, jiang2011, lundkvist2015, campante2017, yu2018, corsaro2020, demoura2020, li2020} to extract the seismic parameters. 
Figure~\ref{fg:powerspectrum} shows the power density spectrum of HD\,76920 computed based on the \texttt{eleanor} light curve, combining both Cycle 1 and Cycle 3 data. The 10-min cadence sectors were rebinned to 30-min cadence, with a simple average over three measurements. The power spectrum shows a frequency-dependent background signal due to stellar activity, granulation, and faculae that can be modeled by a superposition of several Lorentzian-like functions \citep[i.e., Harvey-like models,][]{harvey1985, karoff2008, jiang2011, kallinger2014, corsaro2017}, and a white noise term. The background shown as the cyan dotted curve in Figure~\ref{fg:powerspectrum} was obtained by fitting the background model with two Harvey-like components and one white noise to the smoothed power spectrum. 
The target pixel time series of the Cycle 3 sectors have been detrended with a moving median to remove long timescale stellar variations. As a result, the background noise at low frequency is greatly suppressed during this process. Therefore, the background fit displayed in Figure~\ref{fg:powerspectrum} disregarded frequencies below $\sim$10\,$\muHz$.
The global seismic parameters such as the frequency of maximum power ($\numax$) and the mean large frequency separation ($\Dnu$) were based on the analysis of the background-corrected power spectrum generated by each group. 

\begin{figure}
\resizebox{1.0\hsize}{!}{\includegraphics[]{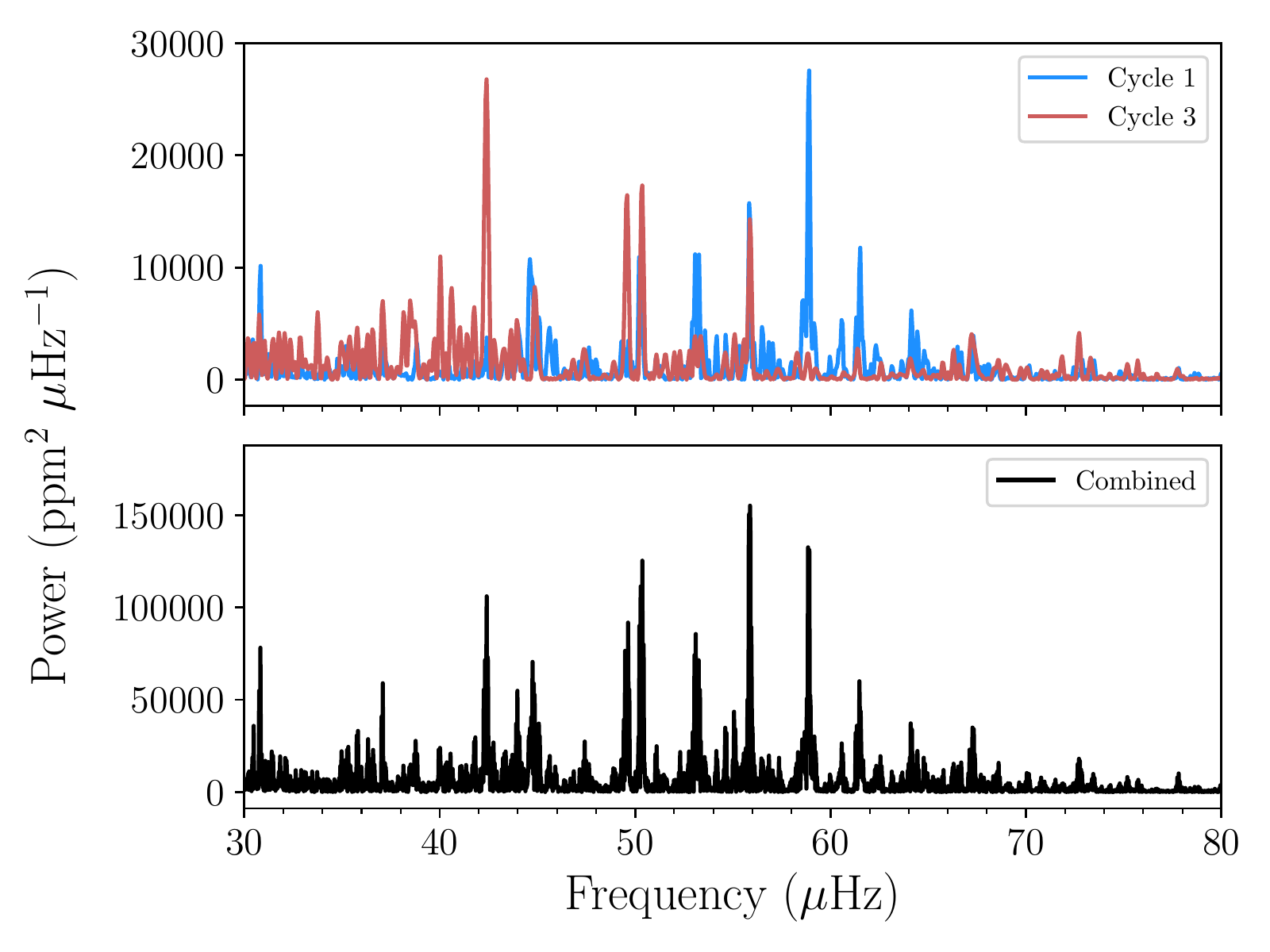}}
\caption{Background corrected power spectra of HD\,76920 depicted in the power excess region. Top panel: power spectra of the Cycle 1 (Sectors 9 to 11) and Cycle 3 (Sectors 36 and 37) light curve. The Cycle 1 data shows a larger value of $\numax$, compared with the Cycle 3 data. Bottom panel: power spectra of the combined light curve that is used for the asteroseismic analysis.}
\label{fg:corrected_ps}
\end{figure}

In general, $\numax$ was measured by fitting a Gaussian distribution profile to the power excess hump of the  smoothed power spectrum \citep[e.g.,][]{hekker2010} or through the 2D autocorrelation function \citep[e.g.,][]{viani2019}. 
To measure $\Dnu$, techniques like autocorrelation of the amplitude spectrum \citep[e.g.,][]{huber2009, mosser2009}, power spectrum of the power spectrum \citep[e.g.,][]{kjeldsen1995, mathur2010, jiang2015}, matched filter response function \citep[e.g.,][]{gilliland2011}, and asymptotic or linear fit to the frequencies of the radial modes (individual mode extraction given in Section~\ref{sc:fre}) were used. 
However, a clear shift of the oscillation power excess region to lower frequencies is observed in the power spectrum generated with Cycle 3 data (Sectors 36 and 37), compared with the Cycle 1 one (Sectors 9 to 11), as illustrated in the upper panel of Figure~\ref{fg:corrected_ps}. 
This is principally due to the stochastic nature of the oscillations so that the change in mode amplitudes impacts the value of $\numax$.
This shift of the power excess region may lead to a larger uncertainty in $\numax$ when compared to the corresponding formal uncertainty originating from different pipelines. 
With this in mind, we computed consolidated results from the 8 independent determinations of the global seismic parameters. In particular, we adopted the mean values of the parameter estimates returned by all methods and recalculated the uncertainties by adding in quadrature the corresponding formal uncertainty and the standard deviation. The consolidated results are $\numax  = 53.2 \pm 2.3\, \muHz$ and $\Dnu = 5.53 \pm 0.15 \,\muHz$. We note that due to the intrinsic change of the distribution of mode amplitudes observed between Cycle 1 and Cycle 3 data, this uncertainty of $\Dnu$ resulting from our statistical consolidation approach, can also be overestimated and can be decreased by examining the different $\Dnu$ values against the \'echelle diagrams \citep{stello2011}. As an alternative, instead of taking into account the scatter across results, we also quote here the pipeline results that are of the smallest deviation from the mean values of all pipeline results as $\numax  = 52.4 \pm 0.3\, \muHz$ and $\Dnu = 5.52 \pm 0.02\,\muHz$, considering both parameters simultaneously (i.e. both parameters returned by the same pipeline). This level of uncertainty of $\Dnu$ is of comparable magnitude to those extracted from 2 sectors of TESS observations of red giants \citep{aguirre2020}.
The combined power spectrum corrected from the background model is shown in the bottom panel of Figure~\ref{fg:corrected_ps}, where signals with $\nu<50\,\muHz$ are largely enhanced by the Cycle 3 data, and those with $\nu>50\,\muHz$ are due to the Cycle 1 data. The asteroseismic analysis discussed in Section~\ref{sc:asteroseismic} were performed based on the combined power spectrum (lower panel of Figure~\ref{fg:corrected_ps}).


\subsection{Individual Mode Frequencies}
\label{sc:fre}

\begin{figure}
\resizebox{1.0\hsize}{!}{\includegraphics[]{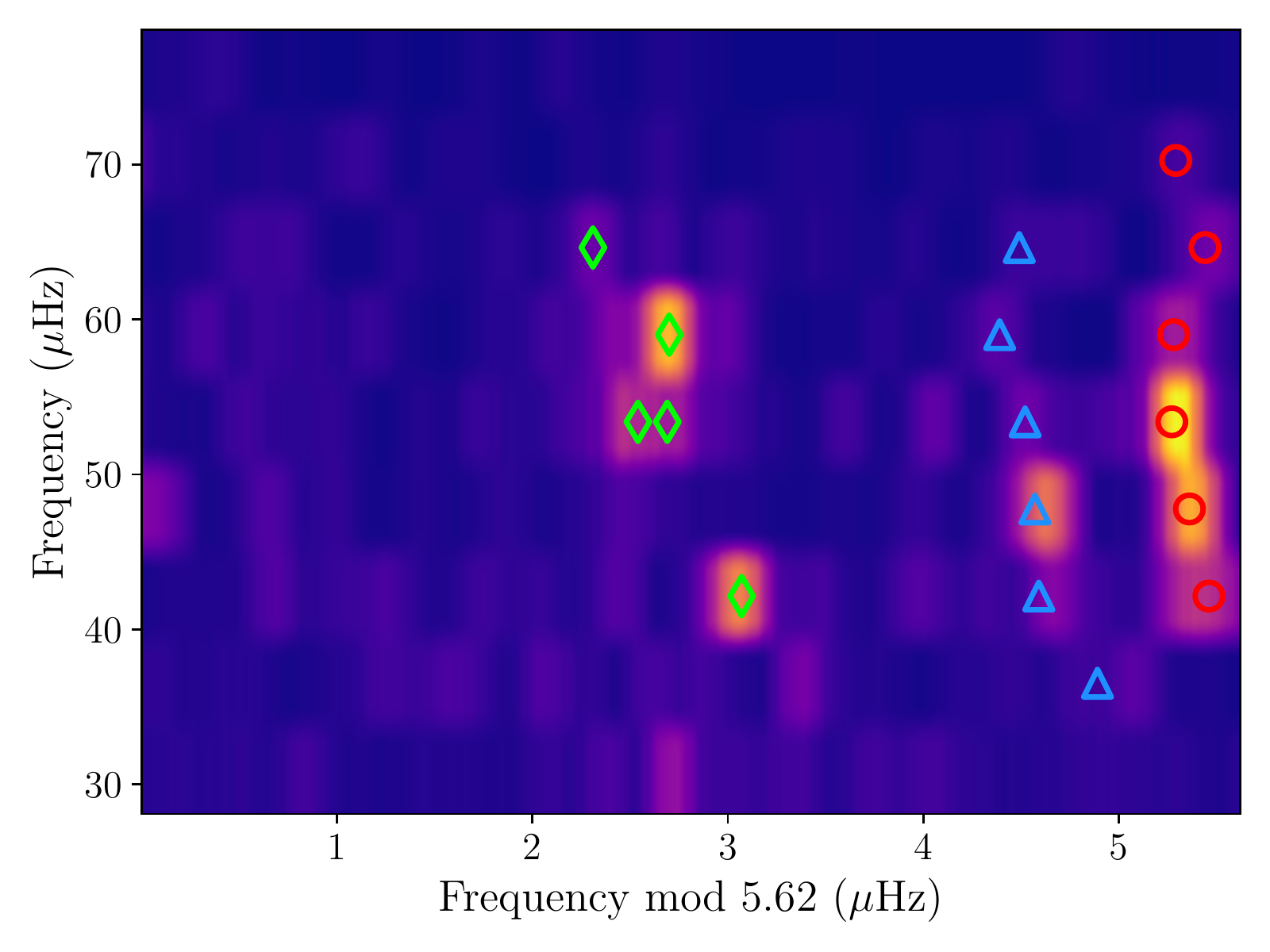}}
\caption{\'Echelle diagram of the background-corrected PSD, folded on an large frequency separation of $\Dnu = 5.62\,\muHz$ estimated by fitting the radial-mode frequencies (c.f. Section~\ref{sc:fre}). Identified individual mode (Table~\ref{tb:frequency}) frequencies are marked with red circles ($\ell = 0$), green diamonds ($\ell = 1$) and blue triangles ($\ell = 2$).}
\label{fg:echelle}
\end{figure}

In the bottom panel of  Figure~\ref{fg:corrected_ps} the combined power spectrum shows a regular series of peaks corresponding to solar-like oscillations within the frequency range between 30 to 80\,$\muHz$. To extract individual oscillation modes from the power spectrum several independent methods ranging from traditional iterative fitting of sine waves, i.e., pre-whitening \citep[e.g.][]{hans05, lenz05, bedding07, jiang2011}, to fitting of Lorentzian mode profiles individually or in a global power spectrum model \citep[e.g.][]{hand11, app12, mosser12, corsaro15, vrard15, davies16, hand17, rox17, kall18, corsaro2020, li2020, breton2022} were used by different pipelines. The extracted mode frequencies and corresponding uncertainties are listed in Table~\ref{tb:frequency}.
The extracted radial modes also allowed us to measure $\Dnu$ by fitting a straight line to the radial-mode frequencies. Thus, the slope of the line is $\Dnu$ that is $5.62 \pm 0.03 \,\muHz$ using this method. 
The \'echelle diagram \citep{bedding2010} generated using this $\Dnu$ is depicted in Figure~\ref{fg:echelle}. The two vertical ridges located near the right edge of the figure correspond to $\ell = 0$ (red circles) and 2 (blue triangles) modes.
However, due to the relatively short duration of our TESS data for HD\,76920, clear mixed-mode pattern is not visible in the power spectrum or in the \'echelle diagram, though there are a few peaks corresponding to $\ell = 1$ mixed modes (green diamonds in Figure~\ref{fg:echelle}) appearing in the spectrum. In Table~\ref{tb:frequency} we also list the $\ell = 1$ modes that are extracted by at least two independent sources, including a pair of mixed modes around $53\,\muHz$ that were simultaneously extracted by 3 independent sources.

\begin{table}[!t]
\begin{center}
\caption{Extracted Oscillation Frequencies from TESS light curve and Mode Identification based on the best-fitting model for HD\,76920}\label{tb:frequency}
\begin{tabular}{lcccc}
\hline
$\ell$  &$n$($n_{\rm p}$) &  $\nu$       &  $\sigma_{\nu}$  & SNR\\
          & & ($\muHz$) &    ($\muHz$) &  \\
\hline
0   &7& 44.80  &  0.13 & 6.63 \\
0   &8& 50.32  &  0.10 & 8.08\\
0   &9& 55.85  &  0.08 & 12.50\\
0   &10& 61.48  &  0.08 & 6.71\\
0   &11& 67.18  &  0.13 & 3.83\\
0   &12& 72.73  &  0.15 & 4.0\\
1 & 6 & 42.41 & 0.08 & 12.05 \\ 
1 & 8 & 53.12 & 0.03 & 9.54 \\
1 & 8 & 53.27 & 0.03 & 6.73 \\
1 & 9 & 58.90 & 0.07 & 6.95 \\
1 &10 & 64.13 & 0.08 & 3.61 \\
2   &5& 38.61  &  0.29 & 3.71 \\
2   &6& 43.93  &  0.24 & 5.88 \\
2   &7& 49.53  &  0.12 & 7.13 \\
2   &8& 55.10  &  0.07 & 5.42 \\
2   &9& 60.59  &  0.20 & 4.48 \\
2   &10& 66.32  &  0.07 & 3.89\\
\hline
\end{tabular}
\end{center}
\tablecomments{Each mode is labeled according to its mode degree $\ell$, radial order $n$ (or radial order of acoustic-component $n_{\rm p}$ for non-radial modes) from the best-fitting model. $\nu$ and $\sigma_\nu$ are the mode cyclic frequency and uncertainty. Modes identified 
simultaneously by at least two independent methods/sources are selected and used as modeling constraints. All these modes have a height-to-background ratio (SNR) larger than 3. The noise at each frequency is calculated as the average amplitude in the residual periodogram in a frequency range ($20\,\muHz$ boxsize) that encloses the mode peak after the frequency is pre-whitened}.
\end{table}

\begin{table}[!t]
\begin{center}
\caption{Modeling configurations from different sources. One entry is used where all five teams used the same input physics.}\label{tb:model}
\begin{tabular}{cccccc}
\hline
Team & Ong & Kayhan & Izmir & Zhang & BESTP \\
\hline
Evolution code & \texttt{MESA} (r12778) & \texttt{MESA} (r12778) &  \texttt{MESA} (r15140) & \texttt{MESA} (r10398) & \texttt{ASTEC} \\
Oscillation code & \texttt{GYRE} &  \texttt{ADIPLS} & \texttt{ADIPLS} &\texttt{ADIPLS} & \texttt{ADIPLS}\\
EoS\tablenotemark{\scriptsize a} & \texttt{MESA}/OPAL &  \texttt{MESA}/OPAL &  \texttt{MESA}/OPAL & \texttt{MESA}/OPAL & OPAL \\
Surface Correction\tablenotemark{\scriptsize b} & BG-2term & KB & KB & None & BG-2term \\
Nuclear reactions & \multicolumn{5}{c}{NACRE \citep{angulo1999}}\\
High-$T$ opacities & \multicolumn{5}{c}{OPAL \citep{iglesias1993, iglesias1996}} \\
Low-$T$ opacities & \multicolumn{5}{c}{\cite{ferguson2005}} \\
Solar mixture\tablenotemark{\scriptsize c} & GS98 & AGSS09 &AGSS09 & GS98 & AGSS09 \\
Atmosphere\tablenotemark{\scriptsize d} & Eddington gray & simple photosphere &simple photosphere & Eddington gray & Kurucz \\
$\alpha_\mathrm{MLT}$ & 1.83 & 2.175 &  1.828  & 2.0 & 1.7 -- 2.1\\
Overshoot & Step overshoot\tablenotemark{\scriptsize e} & None & None & None & None \\
Diffusion & \cite{thoul1994} & None & None & None & None \\
\hline
\end{tabular}
\end{center}
\tablenotetext{\scriptsize a}{\scriptsize The \texttt{MESA}/OPAL tables are based on the 2005 update of the OPAL EoS tables \citep{rogers2002}.}
\tablenotetext{\scriptsize b}{\scriptsize The adopted methods for surface correction are \cite{kjeldsen2008} (KB) and \cite{ball2014} two-term correction (BG-2term)}
\tablenotetext{\scriptsize c}{\scriptsize Solar composition given in \cite{grevesse1998} (GS98) and \cite{asplund2009} (AGSS09) are used for initial chemical composition.}
\tablenotetext{\scriptsize d}{\scriptsize The atmosphere choices used in \texttt{MESA} are introduced in \cite{paxton11},  and \texttt{ASTEC} uses the Kurucz model \citep{kurucz1991} for atmosphere.}
\tablenotetext{\scriptsize e}{\scriptsize The overshoot parameters $f_{\rm ov}$ and $f_{\rm 0, ov}$ are set as 0.006 and 0.003. The overshooting is applied at all convective boundaries.}
\tablerefs{Ong: \citet{mier2017}; Kayhan: \citet{kayhan2019}; Izmir: \citet{yildiz19}; Zhang: \citet{zhang2022}; BESTP: \citet{jiang2021}}
\end{table}

\begin{table}[!t]
\begin{center}
\caption{Modeling results from different sources.}\label{tb:model_results}
\begin{tabular}{cccccc}
\hline
Team & Ong & Kayhan & Izmir & Zhang & BESTP \\
\hline
$M_{\star}$ ($\msun$) & $1.31 \pm 0.06$ & $1.28 \pm 0.16$ & $1.15 \pm 0.14$ & $1.17 \pm 0.01$ & $1.20 \pm 0.10$ \\
$R_{\star}$ ($R_\sun$) & $8.96 \pm 0.14$  & $8.29 \pm 0.14$ & $8.62 \pm 0.39$ & $8.67 \pm 0.02$ & $8.87 \pm 0.26$  \\
$\log g$ (cgs) & $2.650 \pm 0.008$  & $2.709 \pm 0.002$  & $2.630 \pm 0.019$  & $2.630 \pm 0.001$  & $2.620 \pm 0.010$   \\
$t$ (Gyr) & $4.1 \pm 0.7$  & $4.3 \pm 0.5$  & $5.8 \pm 2.1$  & $6.3 \pm 0.06$  & $5.6 \pm 1.4$  \\
\hline
\end{tabular}
\end{center}
\end{table}

\begin{figure}
\resizebox{1.0\hsize}{!}{\includegraphics[]{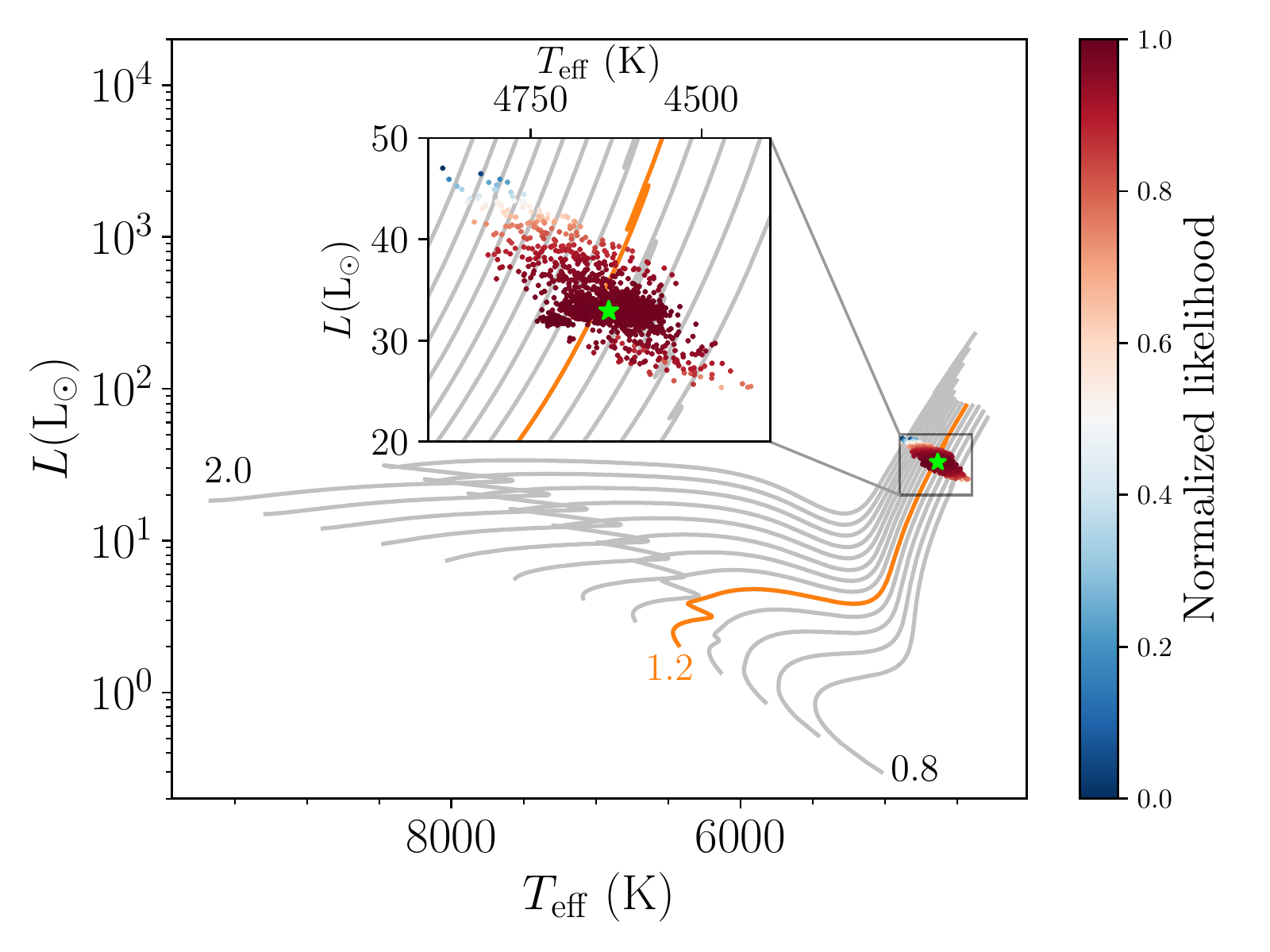}}
\caption{Evolutionary tracks for a series of \texttt{ASTEC} models with different initial masses but same chemical abundance ($X=0.714$ and $Z=0.0142$, corresponding to [Fe/H] = -0.09) and mixing length parameter ($\alpha = 1.927$) that are the optimization outputs of the \texttt{BESTP} pipeline. The initial masses of the models increase from 0.8 to 2.0 $\msun$ with a step of 0.1 $\msun$. The evolutionary track indicated by orange is with an initial mass of 1.2 $\msun$ that is closest to the adopted value (Table~\ref{tb:starpar}).  Sampling points by  \texttt{BESTP} are drawn in the diagram and also in the inset, with color-coded normalized likelihood values. The green star marks the location of our best-fitting model ($M=1.21\,\msun$, $X=0.713$ and $Z=0.0145$) from \texttt{BESTP}.}
\label{fg:hr_diagram}
\end{figure}

\begin{figure}
\resizebox{0.8\hsize}{!}{\includegraphics[]{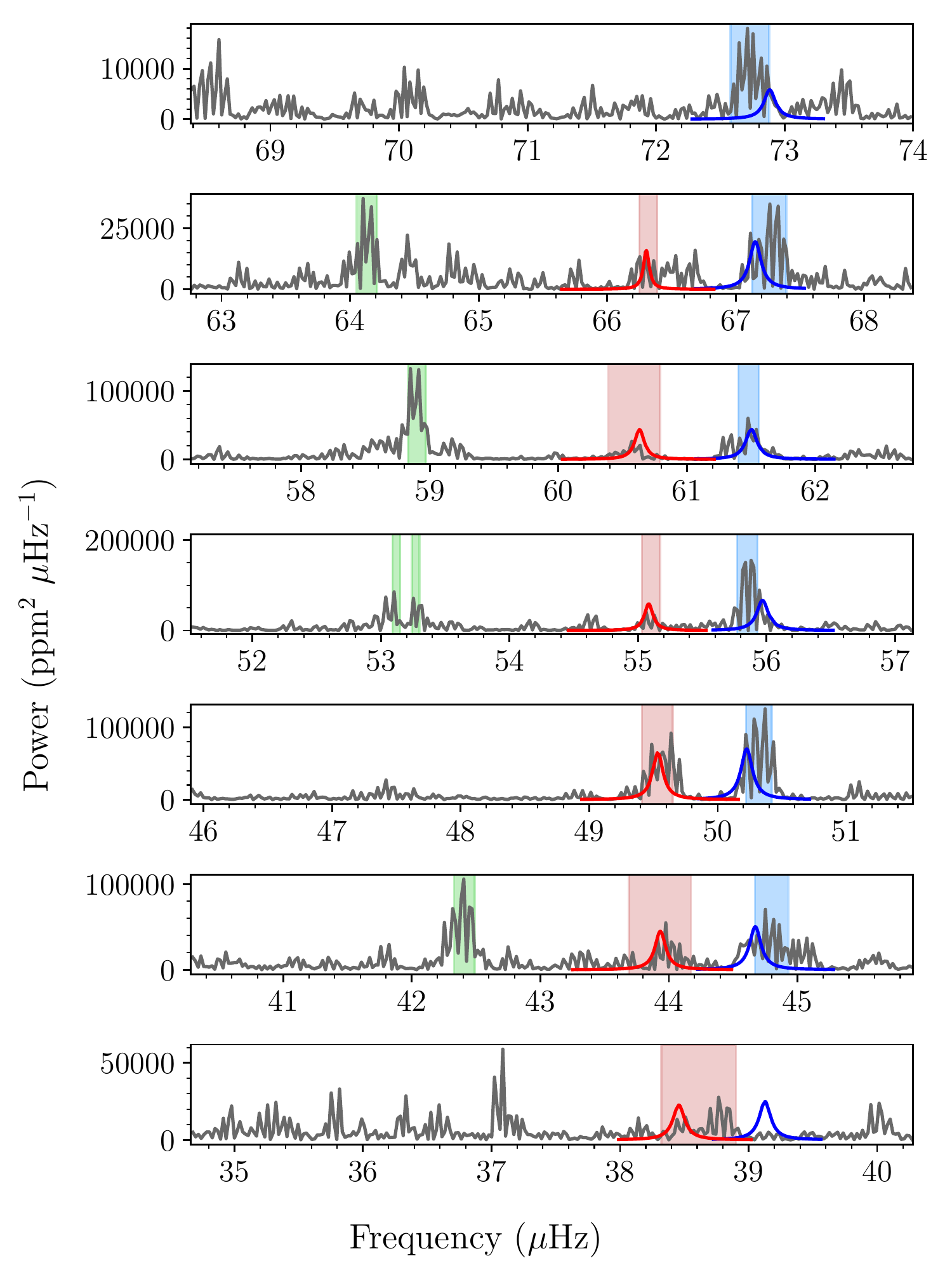}}
\caption{Background-corrected PSD (dark grey) plotted in the \'echelle diagram format that divides the spectrum into bins each $\Dnu$ wide. The blue and red peaks are the $\ell = 0$ and 2 modes, respectively,  of the best-fitting model returned by \texttt{BESTP} (Figure~\ref{fg:hr_diagram}). The line width and amplitude of the theoretical modes are derived using the formulae introduced in \cite{lund17} and  \cite{ball2018}. The corresponding observed frequencies and uncertainties (Table~\ref{tb:frequency}) are indicated by the horizontal spans, with additional $\ell =1$ modes in green.}
\label{fg:compare_ps}
\end{figure}

\section{Asteroseismic Modelling}
\label{sc:model}

Asteroseismic modelling is a powerful tool to estimate fundamental stellar properties. Five independent teams performed modeling efforts to search for the stellar models that best match the classical and asteroseismic constraints from observations for HD\,76920. These teams employ different stellar evolution codes \citep[\texttt{ASTEC, MESA};][]{jcd08a, paxton11, paxton13, paxton15}, oscillation codes \citep[\texttt{ADIPLS, GYRE};][]{jcd08b, town13}, and optimization methods \citep{mier2017, kayhan2019, wu2019, yildiz19, jiang2021, zhang2022}. The input physics adopted by each team is detailed in Table~\ref{tb:model}. Diffusion and overshoot were turned off except for one team. For the modeling, all five teams used the luminosity $L_\star$ of $32.54\pm 1.17\,L_{\sun}$ derived with Gaia DR3 parallax (Section~\ref{sc:spectro}) and the spectroscopic measurements, i.e., the effective temperature $T_{\rm eff}$ and metallicity [Fe/H] from \cite{wittenmyer17}, as constraints. As for the seismic constraints, the large frequency separation $\Dnu = 5.52 \pm 0.02 \, \muHz$ and $\ell = 0$ and 2 mode frequencies in Table~\ref{tb:frequency} were used\footnote{The identified five $\ell = 1$ modes listed in Table~\ref{tb:frequency} do not show clear pattern of mixed modes, thus, provide very limited constraints on the models, as one can always find a match to the observed mode among the dense theoretical dipolar modes.}. 
Generally, the most p-dominated $\ell = 2$ mixed modes of each model are selected to be compared with the observations, except for team Ong who computed the pure quadrupole p-modes of stellar models using the $\pi$-mode isolation condition of \cite{ong2020} as implemented in GYRE\footnote{These pure quadrupole p-modes are matched to the frequencies of the observed p-dominated quadrupole mixed modes, under the assumption that interference from mode mixing in these high-luminosity red giants may be neglected \citep[cf. discussion in][]{ong2021}.}.
The constraint on $\numax$ were not considered in the modeling due to the discrepancy of the oscillation power excess region between the sectors (Section~\ref{sc:global}). We measured the phase shift $\epsilon_{\rm c}$ of the central radial mode, which is the linear offset in the asymptotic fit to the acoustic modes, as $\sim 0.94$, indicating that the star is still on the red giant branch (RGB) burning only hydrogen in a shell, because the more evolved horizontal branch stars (e.g. red clump and secondary clump stars) that have already ignited helium in the core would have a systematically smaller $\epsilon_{\rm c}$ than their counterparts on the RGB \citep{kallinger2012}.  Moreover, horizontal branch tracks hardly cross the observational constraints in the HR diagram, thus, we limited the modeling only to the RGB.

The outputs from the five teams are generally in agreement with each other (Table~\ref{tb:model_results}),
though differences are inevitably seen due to the diversity of modeling codes, procedures and input physics adopted by different teams. We note that the amount of convective overmixing ($f_{\rm ov} = 0.006$) adopted by team Ong is mainly for numerical softening and is too small to induce an appreciable physical difference to the spectroscopic parameters \citep[see discussions in, e.g.,][]{claret2017,claret2018,claret2019,guo2019,zhang2022}. Furthermore, since we have only used even-degree ($\ell = 2$) p-dominated modes in this work, the effects of overshoot on the g-mode cavity \citep[e.g.][]{lindsay2022} are not relevant. For the even-degree p-modes, envelope overshoot will reduce the amplitude of the convective-glitch signature induced into the p-modes, which is already very small on the main sequence, and decreases in amplitude as the convective boundary retreats closer to the core as the star ascends the RGB.
The consolidated values (the mean from all sources) for stellar mass $M_{\star, \rm seis}$, radius $R_{\star, \rm seis}$, age $t$, surface gravity log\,$g$ and density $\rho$ are summarized in Table~\ref{tb:starpar}, constraining the corresponding parameters to a precision level of 9\%, 4\%, 27\%, 1\%, 15\%, respectively, which are the most precise results for HD\,76920 so far. 
The final uncertainties on these stellar parameters were recalculated by adding in quadrature the corresponding formal uncertainty for a given parameter to the standard deviation of the parameter estimates returned by all teams. Therefore, both random and systematic errors arising from the diversity of modeling methods from different teams were taken into account. The estimated log\,$g$ from modeling is distinctly smaller than the spectroscopic results reported by \cite{wittenmyer17}, but matches the one measured by \cite{bergmann21} (who used updated spectroscopy data) within $1\sigma$, confirming the SED results (see Section~\ref{sc:spectro}).

Figure~\ref{fg:hr_diagram} shows the locations of the sampling points by \texttt{BESTP} in the HR diagram, along with a series of evolutionary tracks with different initial masses generated by \texttt{ASTEC}. The sampling points are color-coded according to the normalized likelihood, thus, redder samples have higher possibilities as the representation of the real star. According to the figure, HD\,76920 is most likely approaching closely to the RGB luminosity bump, where the properties of mixed-modes are significantly impacted by the buoyancy glitch \citep{cunha2015, cunha2019, jiang2020b}, a signature that can help us inspect the stellar mid-layer structures \citep{pincon2020, jiang2022}. However, the short observation duration for HD\,76920 limits the detection of mixed-modes from the TESS power spectrum, thus, such investigation of stellar interior structure with the help of mixed modes is not feasible with current TESS data. Nevertheless, by matching the theoretical oscillation frequencies calculated for the best-fitting model with the observed TESS power spectrum (Figure~\ref{fg:compare_ps}), we could confirm the mode degree and identify the mode order for the oscillation modes that were extracted by different groups (Table~\ref{tb:frequency}). 
The theoretical modes used to identify observed modes are corrected for the surface effect \citep[e.g.,][]{houdek2017} that yields a systematic offset between the calculated and the observed oscillation frequencies. For theoretical dipolar mixed modes, correction for the surface effect was not performed due to the lack of enough observed dipolar frequencies, thus, we did not show them in Figure~\ref{fg:compare_ps}.

\section{Characterization of HD\,76920\,\lowercase{b}}

According to the prediction for the transit ephemeris made by \cite{bergmann21}, if HD\,76920\,b were to transit, it would have done so during TESS Sector 9.
However, no clear transit signal was found by \cite{bergmann21} or in our data. Thus, we could not measure the planetary orbit or eccentricity through the TESS photometric data.

However, we can combine our newly computed asteroseismic stellar mass ($M_{\star, \rm seis} =  1.22 \pm 0.11\, \msun$) with the orbital period, RV semi-amplitude and eccentricity values found by \cite{bergmann21} to compute updated values for the planet's semi-major axis and minimum mass. Using values from \cite{bergmann21}, we find $a=1.165 \pm 0.035$~au and $M_{\rm p}\sin{i} = 3.57 \pm 0.22\,M_{\rm Jup}$.
These updated parameters place the orbital pericenter of the planet at $0.142 \pm 0.005$~au, revealing that the planet's orbit is sufficiently far from the star such that tidal decay is currently negligible until the star expands its radius by at least 50-100\% \citep{villaver2009} and subsequently engulfs the planet, in line with the findings of \cite{bergmann21}.

The time at which the star will engulf the planet will depend on when the star began ascending the RGB. During this phase of stellar evolution, the star's radius will expand to a distance approximately one order of magnitude greater than the planet's orbital pericenter, allowing the star to easily and quickly engulf the planet. For instance, further evolving the best-fitting model (Figure~\ref{fg:hr_diagram}) along the evolutionary track, we predict that the expanding star will have a radius one order of magnitude greater than the planet's orbital pericenter and engulf the planet in about 130\,Myr.

As an independent estimation of the planetary engulfment time, we also calculated models using the {\tt SSE} prescription from \cite{hurley2000}.  According to this prescription, a $1.22\,M_{\odot}$ star with a metallicity of $Z=0.02$ will begin the red giant phase at about 5.58\,Gyr and leave it at 6.01\,Gyr.
During this interval, the planet will be engulfed. Asteroseismology has allowed us to estimate the current age of the star to be $5.2 \pm 1.4$\,Gyr. This value clearly demonstrates that the star is on the verge of engulfing the planet. However, the uncertainty on the age is about triple the value of the duration of the red giant phase. Nevertheless, asteroseismology has also allowed us to constrain the stellar radius to a value of $8.68 \pm 0.34\,R_{\odot}$.  This value corresponds to an age from the {\tt SSE} model of $5.939 \pm 0.004$\,Gyr, meaning that the planet will be engulfed in less than about 150\,Myr. However, this approximate upper bound is likely an overestimate given the fact that the star's metallicity is sub-Solar.

Other stellar models produce similar results, just renormalized to other absolute ages within the measured asteroseismic uncertainty. For example, a $1.22\,M_{\odot}$ star with a sub-Solar metallicity of $Z=0.01$ will begin the red giant phase at about 4.72\,Gyr and leave it at 5.08\,Gyr. In this case, the asteroseismically measured radius corresponds to an age from the {\tt SSE} model of $5.007 \pm 0.004$\,Gyr. This value illustrates that the planet will be engulfed within about 70\,Myr. This result is in agreement with that of \cite{bergmann21}.

\section{Conclusions}

In this work, we have analyzed the TESS photometric data for HD\,76920 to determine the star's fundamental parameters using asteroseismology, and to characterize the exoplanet system consisting of a planet with an extremely large orbital eccentricity. In total, 5 sectors of TESS light curves are used for the extraction of asteroseismic parameters for the host star, including 17 individual oscillation frequencies. Modeling by various pipelines that utilize the extracted asteroseismic parameters, classical spectroscopic observables, as well as luminosity from Gaia DR3 parallax, places strong constraints on the stellar parameters. Through the asteroseismic analysis, we obtain a value for the stellar mass of $1.22 \pm 0.11\, \msun$, a stellar radius of $8.68 \pm 0.34\,R_\sun$ and an age of  $5.2 \pm 1.4$\,Gyr, which provide the most precise estimations for HD\,76920 to date. The stellar models reveal that the star is ascending the red giant branch and most likely approaching closely to the luminosity bump where the properties of mixed-modes are significantly impacted by the buoyancy glitch. However, the current power spectrum of HD\,76920 from TESS does not allow the extraction of a sufficient number of mixed modes as would be required for the investigation of the buoyancy glitch.

The updated stellar parameters of the host star from our asteroseismic analysis have enabled improved estimations for the semi-major axis and mass of the planet as $a=1.165 \pm 0.035$\,au and $M_{\rm p}\sin{i} = 3.57 \pm 0.22 \, M_{\rm Jup}$. With an orbital pericenter of $0.142 \pm 0.005$\,au, we confirm that the planet is currently far away enough from the star to experience negligible tidal decay before being engulfed in the stellar envelope. However, we predict that this event will occur within about 100\,Myr, depending on the stellar model used.

HD\,76920 will be observed in 5 more sectors by TESS in Cycle 5. The prolonged data will possibly enable the detection of mixed-modes, and thus, the investigation of stellar interior through these modes. Moreover, our asteroseismic analysis emphasizes the potential of TESS for characterizing exoplanet systems through the synergy between exoplanet research and asteroseismology.

\section{acknowledgments}
The project leading to this publication has received funding from the B-type Strategic Priority Program of the Chinese Academy of Sciences (Grant No. XDB41000000). ADF acknowledges the support from the National Science Foundation Graduate Research Fellowship Program under Grant No.\,(DGE-1746045). M.S.L would like to acknowledge the support from VILLUM FONDEN (research grant 42101) and The Independent Research Fund Denmark's Inge Lehmann program (grant agreement no.:  1131-00014B). CK is supported by Erciyes University Scientific Research Projects Coordination Unit under grant number DOSAP MAP-2020-9749. T. Wu thanks the supports from the National Key Research and Development Program of China  (Grant No. 2021YFA1600402), from the National Natural Science Foundation of China (Grant Nos.  11873084, 12133011, and 12273104), from Youth Innovation Promotion Association of Chinese Academy of Sciences, and from Ten Thousand Talents Program of Yunnan for Top-notch Young Talents. X.-Y. Zhang thanks the support from the National Natural Science Foundation of China (Grant No. 12173105). S.M.~acknowledges support from the Spanish Ministry of Science and Innovation (MICINN) with the Ram\'on y Cajal fellowship no.~RYC-2015-17697, grant no.~PID2019-107187GB-I00 and PID2019-107061GB-C66, and through AEI under the Severo Ochoa Centres of Excellence Program 2020--2023 (CEX2019-000920-S). R.A.G. acknowledges the support from PLATO and GOLF CNES grants. This work was also supported by Funda\c c\~ao para a Ci\^encia e a Tecnologia (FCT) through research grants UIDB/04434/2020 and UIDP/04434/2020. TLC and MSC are supported by FCT in the form of a work contracts (CEECIND/00476/2018, CEECIND/02619/2017). D.H. acknowledges support from the Alfred P. Sloan Foundation and the National Aeronautics and Space Administration (80NSSC21K0652, 80NSSC20K0593).

\facilities{TESS, Gaia}

\software{\texttt{eleanor} \citep{2019PASP..131i4502F},
          \texttt{lightkurve} \citep{2018ascl.soft12013L},
          \texttt{TessCut} \citep{2019ascl.soft05007B}},
          \texttt{LAURA} \citep{demoura2020}, \texttt{FAMED} \citep{corsaro2020},
          \texttt{echelle}, \texttt{SolarlikePeakbagging} \citep{li2020}, \texttt{apollinaire} \citep{breton2022},
          \texttt{fnpeaks},  \texttt{SYD} \citep{huber2009}, \texttt{A2Z} \citep{mathur2010}, 
          \texttt{Period04} \citep{lenz05},
          \texttt{Yabox} \citep{mier2017}, \texttt{SEDEX} \citep{yu2021, yu2022}

\bibliography{draft}{}
\bibliographystyle{aasjournal}
\clearpage

\end{CJK*}
\end{document}